\documentclass[prl,reprint,superscriptaddress,aps]{revtex4-1}

\usepackage[utf8]{inputenc}
\usepackage{graphicx,xcolor}

\usepackage{hyperref}

\definecolor{dark-blue}{rgb}{0,0.2,0.6}

\hypersetup{%
    unicode=true,%
    colorlinks=true,%
    linkcolor=dark-blue,%
    citecolor=dark-blue,%
    urlcolor=dark-blue,%
    pdftitle={Clock-line photoassociation of strongly bound dimers in a magic-wavelength lattice},%
    pdfauthor={O. Bettermann, N. Darkwah Oppong, G. Pasqualetti, L. Riegger, I. Bloch, and S. Fölling},%
    pdfproducer={},%
    pdfcreator={}}

\usepackage{etoolbox}
\makeatletter
\pretocmd{\NAT@open}{\begingroup\color{\@citecolor}}{}{}
\apptocmd{\NAT@close}{\endgroup}{}{}
\makeatother

\usepackage{amsmath}
\usepackage{textcomp}

\newcommand{\ket}[1]{\ensuremath{\left|{#1}\right\rangle}}

\newcommand{\braket}[2]{\ensuremath{\left\langle{#1}|{#2}\right\rangle}}

\newcommand{\yb}[1]{\ensuremath{{^{\text{#1}}\text{Yb}}}}

\newcommand{\tP}[1]{\ensuremath{{^3\mathrm{P}_{#1}}}}

\newcommand{\sS}[1]{\ensuremath{{^1\mathrm{S}_{#1}}}}

\newcommand{\asciimathunit}[1]{\ensuremath{\,\mathrm{#1}}}

\newcommand{\Erec}{E_\mathrm{rec}}

\newcommand{\ms}{\asciimathunit{ms}}

\newcommand{\nm}{\asciimathunit{nm}}

\newcommand{\kHz}{\asciimathunit{kHz}}

\newcommand{\THz}{\asciimathunit{THz}}
\newcommand{\Gauss}{\asciimathunit{G}}

\newcommand{\subfigref}[2]{\hyperref[fig:#1]{\ref*{fig:#1}(#2)}}
\newcommand{\subfigrefs}[3]{\hyperref[fig:#1]{\ref*{fig:#1}(#2)--(#3)}}
\newcommand{\cfigref}[2]{\hyperref[fig:#1]{\ref*{fig:#1}#2}}

\begin{document}



\title{Clock-line photoassociation of strongly bound dimers in a magic-wavelength lattice}

\author{O.~Bettermann}
\email{Oscar.Bettermann@physik.uni-muenchen.de}
\author{N.~\surname{Darkwah~Oppong}}
\thanks{These authors contributed equally to this work}
\author{G.~Pasqualetti}
\thanks{These authors contributed equally to this work}
\author{L.~Riegger}
\author{I.~Bloch}
\author{S.~F{\"o}lling}
\affiliation{Ludwig-Maximilians-Universit{\"a}t, Schellingstra{\ss}e 4, 80799 M{\"u}nchen, Germany}
\affiliation{Max-Planck-Institut f{\"u}r Quantenoptik, Hans-Kopfermann-Stra{\ss}e 1, 85748 Garching, Germany}
\affiliation{Munich Center for Quantum Science and Technology (MCQST), Schellingstra{\ss}e 4, 80799 M{\"u}nchen, Germany}
\date{\today}


\begin{abstract}

We report on the direct optical production and spectroscopy of \sS0\mbox{--}\tP0 molecules with large binding energy using the clock transition of \yb{171}, and on the observation of the associated orbital Feshbach resonance near $1300\Gauss$.
We measure the magnetic field dependence of the closed-channel dimer and of the open-channel pair state energy via clock-line spectroscopy in a deep optical lattice.
In addition, we show that the free-to-bound transition into the dimer can be made first-order insensitive to the trap depth by choice of the lattice wavelength.
Finally, we determine the fundamental intra- and interorbital scattering lengths and probe the stability of the corresponding pair states, finding long lifetimes in both interorbital interaction channels.
These results are promising both for molecular clocks and for the preparation of strongly-interacting multiorbital Fermi gases.

\end{abstract}

\maketitle


%
%
%
%
%
Alkaline-earth(-like) atoms (AEAs) such as ytterbium and strontium have attracted considerable interest in the field of ultracold atoms~\cite{he:2019} due to their richer level structure compared to alkali metals.
AEAs feature a long-lived \tP0 state ($\ket{e}$) connected to the \sS0 ground state ($\ket{g}$) via an ultra-narrow clock transition, which is harnessed in today's most precise and accurate optical lattice clocks~\cite{ludlow:2015, takano:2016, mcgrew:2018, bothwell:2019}.
Moreover, the clock state provides an independent degree of freedom for quantum simulations of multiorbital many-body physics~\cite{gorshkov:2010, foss-feig:2010, foss-feig:2010b, silva-valencia:2012, riegger:2018, kanasz-nagy:2018, nakagawa:2018, darkwahoppong:2019, goto:2019, zhang:2020, sotnikov:2020}.

At the low energies and densities in these systems, interactions between atomic pairs are governed by their molecular interaction potential.
In this regime, the properties of the bound states supported by this potential close to the dissociation energy threshold determine the scattering properties.
In addition, if the energy of such a bound state can be suitably adjusted via a Zeeman shift, the interaction strength becomes precisely tunable with an external magnetic field.
Such a mechanism lies at the heart of the magnetic Feshbach resonances utilized in alkali atoms~\cite{chin:2010}.
In the case of an interorbital interaction potential, this leads to orbital Feshbach resonances (OFR)~\cite{zhang:2015}, until now only observed in $\yb{173}$~\cite{hoefer:2015, pagano:2015}.
This novel type of Feshbach resonance allows to tune contact interactions between $\ket{g}$ and $\ket{e}$ atoms, and has enabled the realization of multiorbital Fermi polarons~\cite{darkwahoppong:2019}, coherent preparation of weakly bound dimers on the clock transition~\cite{cappellini:2019}, and has inspired multiple proposals for the realization of exotic superfluidity~\cite{iskin:2016,zou:2018,laird:2020}.

The dimer state associated with the OFR in \yb{173} is weakly bound and intrinsically in the resonant regime, since its binding energy is comparable to other energy scales such as the band structure, the Fermi energy, or typical temperatures.
Here, we report on a strongly bound molecular state in \yb{171}, outside the purely universal regime~\cite{ferlaino:2009}.
The binding energy of this molecule, while still close to the regime of a universal halo dimer, far exceeds all typical energy scales in cold atom ensembles.
In particular, this also applies to the level spacing of the optical lattice traps typically used when driving the clock transition.
The molecular wavefunction is therefore largely independent of the trap parameters, enabling a photoassociation process on a unique and particularly well-defined optical transition.
This is especially promising in the context of molecular clocks, which have been proposed as sensitive probes for possible variations of fundamental constants and gravitation on microscopic scales~\cite{zelevinsky:2008,kotochigova:2009,borkowski:2018,kondov:2019}.
Additionally, the OFR associated with the bound state in \yb{171} occurs at a large magnetic field where spin-exchange interactions are strongly suppressed, in contrast to the OFR in \yb{173}~\cite{hoefer:2015}.

We directly produce interorbital dimers in \yb{171} by addressing the narrow transition from pairs of weakly interacting $\ket{g}$ atoms to the least-bound molecular state in a deep three-dimensional (3D) optical lattice.
First, we measure the energy of the dimer and of the repulsively interacting pair state at magnetic fields up to $1600\Gauss$.
We find a large dimer binding energy of $h\times 292.1(2)\kHz$ at zero magnetic field, in strong contrast with the very shallow bound state in \yb{173}~\cite{hoefer:2015}, and the largest binding energies previously observed with photoassociation on doubly-forbidden transitions~\cite{kato:2013}.
Our results are furthermore well-described by a basic single-channel model corresponding to an OFR at $1300(44)\Gauss$. 
In a second experiment, we show that the free-to-bound transition can be made first-order insensitive to the trap depth by adjusting the wavelength of the optical lattice, an essential step for the implementation of molecular clocks with this state.
Furthermore, we precisely extract the intra- and interorbital scattering lengths via clock-line spectroscopy at small magnetic fields, finding values similar to the ones reported in Ref.~\cite{ono:2019}.
Finally, we probe the stability of the corresponding eigenstates and find long-lived interorbital pair states, which are crucial for the implementation of many-body physics.

%
%
The characterization of the dimer and OFR in \yb{171} requires a detailed study of the magnetic-field-dependent interaction between two atoms in distinct orbitals ($\ket{g}$ or $\ket{e}$) and nuclear spin states ($m_F=\pm1/2$, denoted by $\ket{\uparrow}$ and $\ket{\downarrow}$).
At large magnetic fields $B$, the Hamiltonian is dominated by the differential Zeeman shift $|\delta(B)|= h\times 399.0(1)\,\mathrm{Hz}/\mathrm{G}\times B/2$~\cite{SM} between $\ket{g\uparrow}$ and $\ket{e\uparrow}$ (or $\ket{g\downarrow}$ and $\ket{e\downarrow}$).
At typical interparticle distances, the energetically accessible open channel $\ket{o}$ and inaccessible closed channel $\ket{c}$ for the two-particle scattering problem are then given by $\ket{eg\uparrow\downarrow}=\left(\ket{e\uparrow}\ket{g\downarrow}-\ket{g\downarrow}\ket{e\uparrow}\right)/\sqrt{2}$ and $\ket{eg\downarrow\uparrow}=\left(\ket{e\downarrow}\ket{g\uparrow}-\ket{g\uparrow}\ket{e\downarrow}\right)/\sqrt{2}$, respectively.
A~free atomic pair entering $\ket{o}$ can couple to the least-bound state $\ket{b_c}$ supported by the closed-channel interatomic potential.
This leads to an OFR when the corresponding molecular binding energy is offset by the differential Zeeman shift to the open channel, which changes the entrance energy by $\delta(B)$~\cite{zhang:2015}.

%
%
In our experiment, we prepare a degenerate spin-balanced Fermi gas of \yb{171} in $\ket{g\downarrow}$ and $\ket{g\uparrow}$.
We adiabatically load the atoms into a $30\,\Erec$ deep and nearly isotropic 3D optical lattice operating at the magic wavelength $\lambda_m=759.4\nm$~\cite{lemke:2009}, where atoms in $\ket{g}$ and $\ket{e}$ experience the same trapping potential.
Here, $\Erec = h \times 2.0\kHz$ is the recoil energy of a lattice photon.
The temperature is sufficiently low such that all atoms are in the motional ground state of each lattice site and $\approx25\%$ of the populated sites are doubly occupied.

\begin{figure}[t]
  \includegraphics[width=\columnwidth]{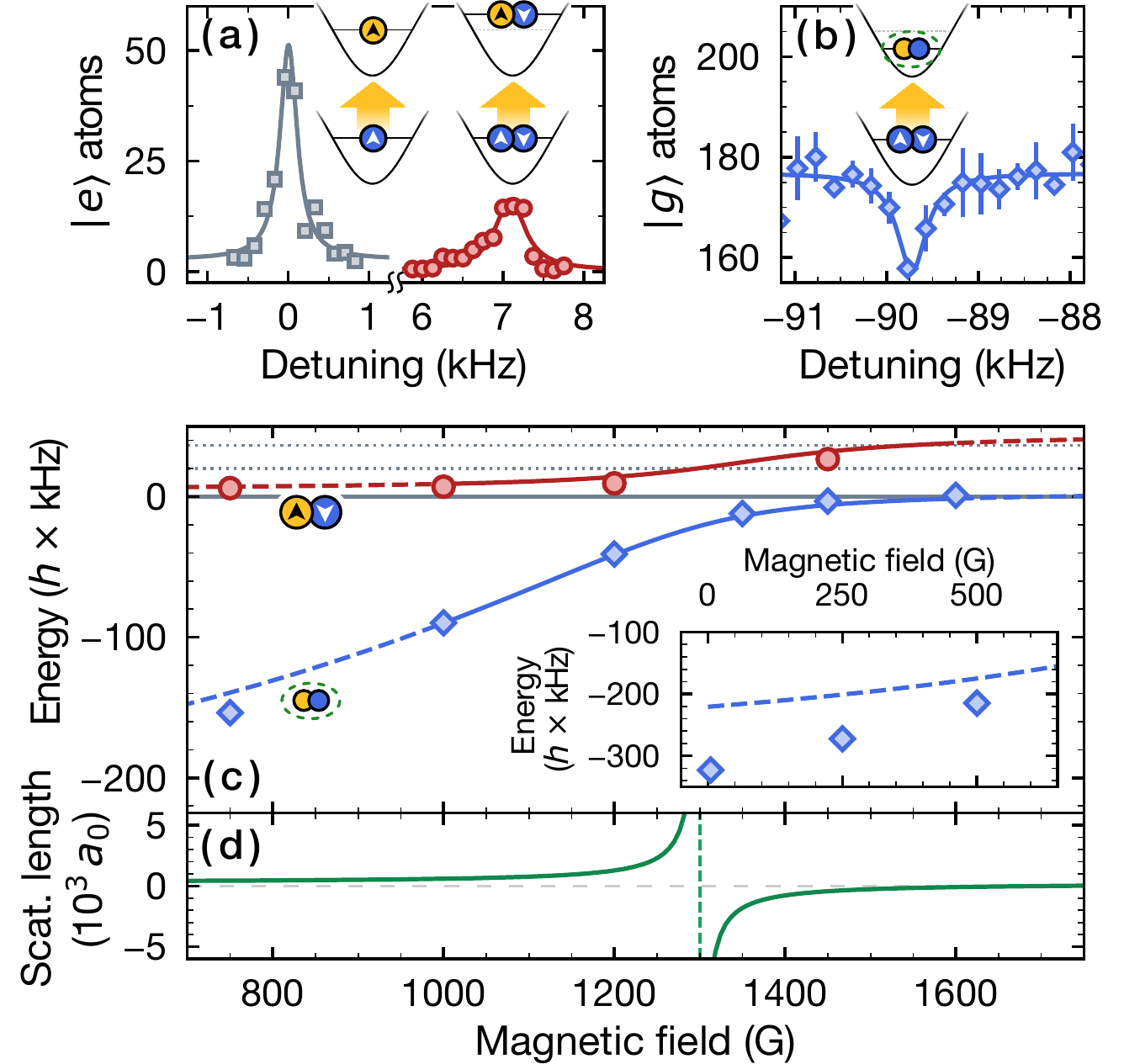}
  \caption{\label{fig:feshbach}
    Clock-line spectroscopy across the orbital Feshbach resonance.    
    \mbox{(a)--(b)}~Raw spectra at $1000\Gauss$ for the single-particle (gray squares), open-channel pair (red circles) and dimer (blue diamond) states.
    We use either the detection of $\ket{e}$ or $\ket{g}$ atoms to maximize the signal.
    The solid lines refer to Lorentzian fits and the error bars in (b) correspond to the standard error of two measurements.
	Blue (yellow) circles in the inset represent $\ket{g}$ ($\ket{e}$) atoms.
    (c)~Energy of the open-channel pair state (red circles) and closed-channel dimer (blue diamonds) relative to the single-particle energy (solid gray line) at variable magnetic field.
    Data coinciding with the first or second band excitation (dotted gray lines) is not shown.    
    In the inset, we show the dimer transition energy at small fields.
   We fit the data to our theoretical model between $1000$ and $1600\Gauss$ (solid lines), where the universal Feshbach relations hold well for the bound state.  
    At other fields, we show the model as dashed lines.    
    (d)~Scattering lengths (solid line) and resonance position (dashed line) extracted from the fit in (c).
  }
\end{figure}

%
%
We measure the transition energy to the dimer $\ket{b_c}$ and to the open-channel pair state $\ket{o}$ (which approaches the band excitation of the non-interacting system at large magnetic fields) for magnetic fields up to $1600\Gauss$, as shown in Fig.~\subfigrefs{feshbach}{a}{c}.
We directly drive on-site $\ket{gg}$ pairs to the desired interorbital state via the clock transition and measure the line shift compared to the $\ket{g\uparrow}\rightarrow\ket{e\uparrow}$ single-particle transition.
Where the transition energies are above $-h \times 100\kHz$, we drive the dimer transition with long, low-intensity pulses while we use short, high-intensity pulses for lower energies.
The open-channel state, and similarly the single-atom reference transition, are addressed with $5\ms$ coherent (Fourier-limited) pulses.
The laser driving the clock transition is linearly polarized and propagating along one of the lattice axes.
To reduce the effects of spatial inhomogeneity, we consider only a small region in the center of the atomic cloud, where $\approx40\%$ of the atoms are on doubly-occupied lattice sites~\cite{SM}.
We obtain a transition energy of $-h\times323.9(2)\kHz$ at zero magnetic field, linearly extrapolating from the 1.1G data point.
We therefore obtain a molecular binding energy of $\epsilon_b=h\times292.1(2)\kHz,$ taking into account the difference in the harmonic oscillator ground state energies of the initial two-particle and final molecular states.
With a separate measurement in pancake-shaped traps, we verify the production of the same molecules in a bulk gas~\cite{SM}.
The binding energy is much larger than all other energy scales in the system such as the temperature, the Fermi energy or the lattice band gap, in stark contrast to the case of the near-resonant bound state in \yb{173}~\cite{hoefer:2015}.
The dimer is therefore only weakly affected by the external confinement owing to the small size of its wavefunction.
In the vicinity of the expected position of the OFR, between $1200$ and $1450\Gauss$, we observe a strong loss of contrast on the transition to $\ket{o}$.
At large fields above $1300\Gauss$, the binding energy vanishes as it approaches the threshold, a result of the diverging scattering length due to the OFR [see Fig.~\subfigref{feshbach}{d}].

To describe the above transition energies, we approximate each optical lattice site as a harmonic trap, allowing us to apply an analytic solution for the interacting atom pairs in our system~\cite{busch:1998}.
We account for the lattice anharmonicity using first-order perturbation theory, in agreement with results from exact diagonalization in our regime~\cite{deuretzbacher:2008,cappellini:2014,SM}.
Around the resonance, we use this model with the simple expression $a(B)=a_{bg}\left[1-\Delta/(B-B_0)\right]$ for the scattering length~\cite{chin:2010}.
From this, we extract the position $B_0=1300(44)\Gauss$, the width $\Delta=402(169)\Gauss$, and the background scattering length $a_{bg}=255(24)\,a_0$, where $a_0$ denotes the Bohr radius.

%
%
The deeply-bound dimer produced in the above measurement is a promising candidate for the implementation of optical molecular clocks.
On each site of the optical lattice, however, the wavefunction of the molecule samples a smaller region of the trapping potential compared to the free atom pair, modifying the free-to-bound transition energy depending on the local trap depth.
This effectively broadens the molecular line due to the inhomogeneous on-site trapping potential in the experiment.
We compensate this effect by introducing a suitable differential ac Stark shift between $\ket{g}$ and $\ket{e}$.
This is achieved to first order by detuning the optical lattice from the magic wavelength of the single-particle transition.
We note that the obtained wavelength is not universal, but specific to the selected magnetic field and lattice depth.

\begin{figure}[t!]
  \includegraphics[width=\columnwidth]{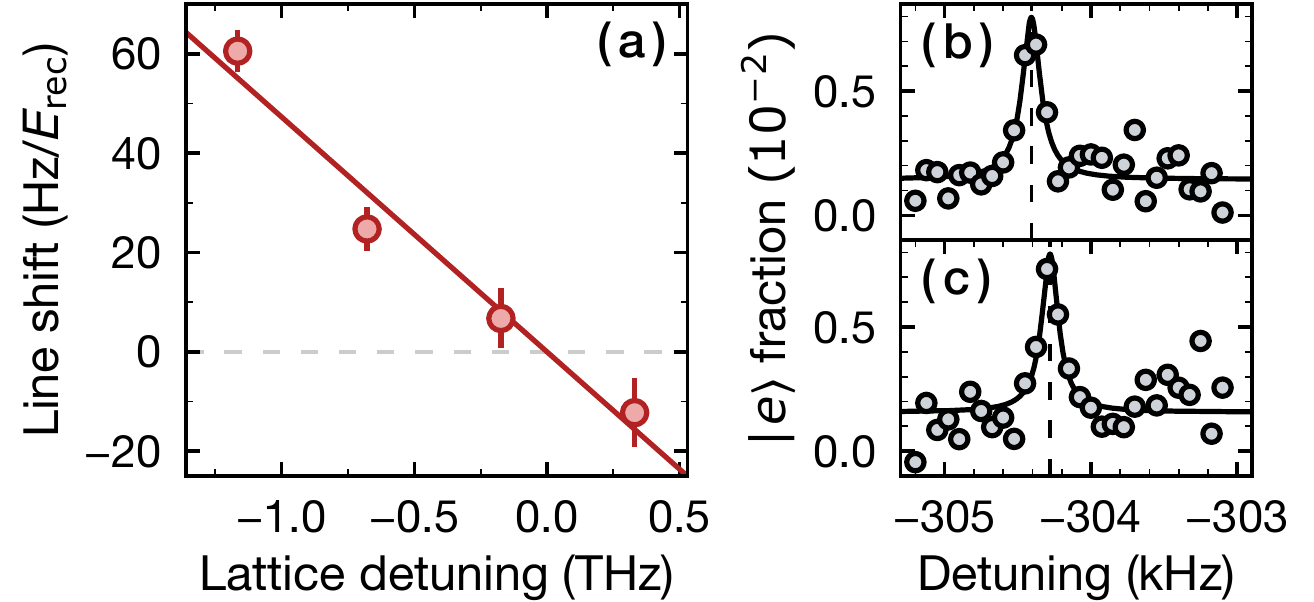}
  \caption{\label{fig:magic_wavelength}
    Lattice-depth dependence of the free-to-bound transition at $1.1\Gauss$.
    (a)~Linear shift of the transition relative to the lattice depth for varying detuning of the lattice laser.
    Each shift is determined from two consecutive measurements in a $25$ and $30\,E_\mathrm{rec}$ deep lattice, with $E_\mathrm{rec}$ the photon recoil energy at the given lattice wavelength.
    The solid line corresponds to a linear fit and error bars denote the fit uncertainty of individual line shapes.
    The detunings are shown relative to the wavelength $776.6(3)\nm$ [$386.0(1)\THz$], which yields the first-order insensitive transition.   
    \mbox{(b)--(c)}~Clock-line spectroscopy of the free-to-bound transition at $778\nm$ ($385.3\THz$) in a (b)~$25.79(7)$ and (c)~$30.95(8)\,E_\mathrm{rec}$ deep lattice. 
    The circles correspond to the fraction of $\ket{e}$ atoms and the solid line denotes a Lorentzian fit.   
    We calculate the shift in (a) from both resonance positions (dashed lines).}
\end{figure}

We explore this approach at a magnetic field $B=1.1\Gauss$ by measuring how the free-to-bound transition line shifts between lattice depths of $25$ and $30\Erec$ for a range of lattice wavelengths $\lambda_{\mathrm{lat}}$ from $776$ to $779\nm$ and observe a linear dependence of the line shift on $\lambda_{\mathrm{lat}}$.
We find that the transition frequency becomes independent of the lattice depth and the molecular line the narrowest for $\lambda_\mathrm{lat}=776.6(3)\nm$, as shown in Fig.~\ref{fig:magic_wavelength}.
%
%
%
In the limit of zero magnetic field, the two-particle Hamiltonian is interaction-dominated and its eigenstates are given by $\ket{eg^+}=(\ket{eg\uparrow\downarrow}-\ket{eg\downarrow\uparrow})/\sqrt{2}=\left(\ket{eg}+\ket{ge}\right)/\sqrt{2}\,\otimes\ket{s}$ and $\ket{eg^-}=(\ket{eg\uparrow\downarrow}+\ket{eg\downarrow\uparrow})/\sqrt{2}=\left(\ket{eg}-\ket{ge}\right)/\sqrt{2}\,\otimes\ket{t}$, connecting to $\ket{o}$ and $\ket{c}$, respectively.
Here, $\ket{s}$ denotes the nuclear spin singlet and $\ket{t}$ the nuclear spin triplet.
In an optical lattice, the on-site interaction energies $U_{eg}^\pm$ directly depend on the corresponding scattering lengths $a_{eg}^\pm$, which contain fundamental information about the interorbital interactions of an $\ket{eg}$ pair.

\begin{figure}[t!]
  \includegraphics[width=\columnwidth]{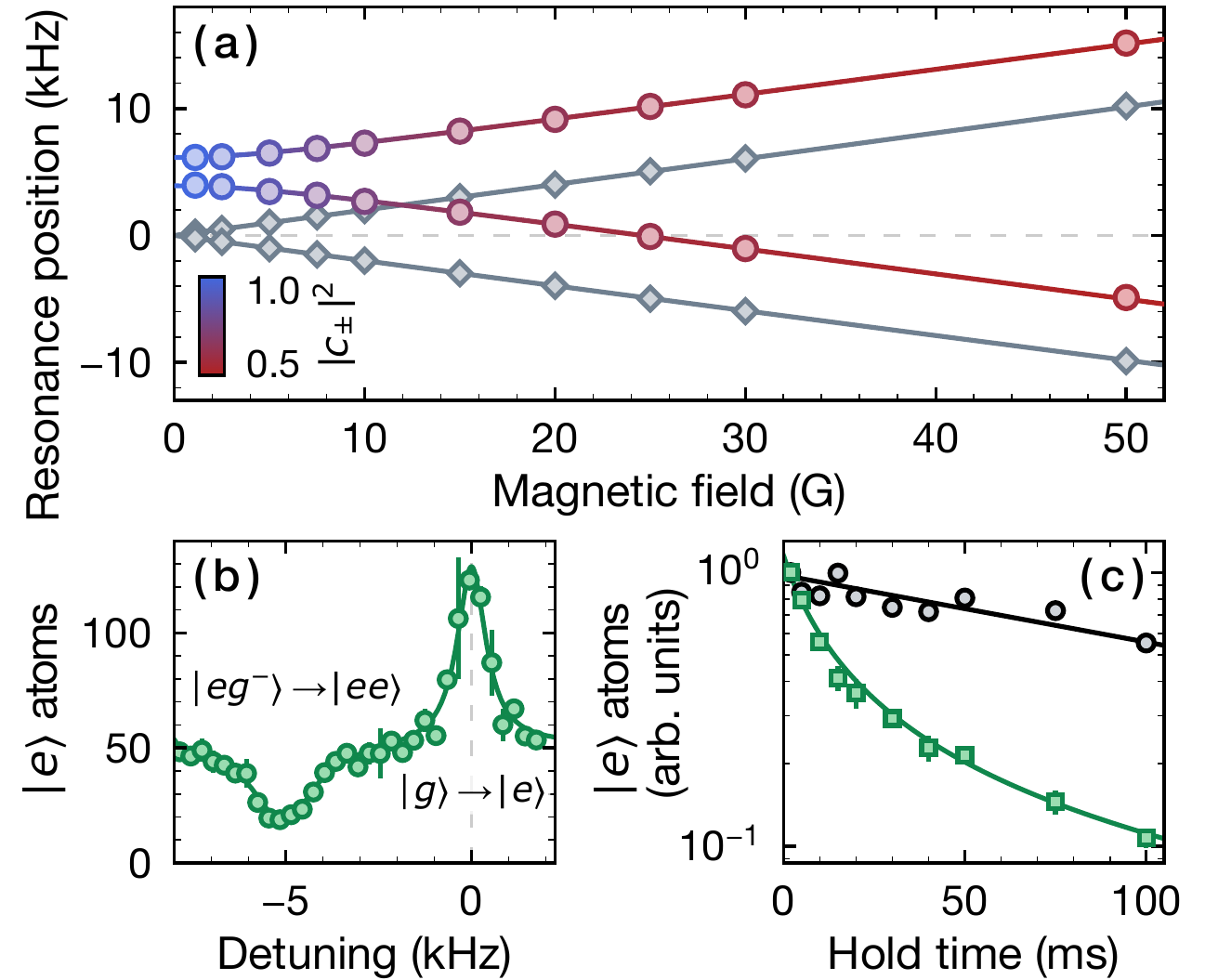}
  \caption{\label{fig:int_spectroscopy}
    Spectroscopy of pair states at low magnetic fields.
    (a)~Transition energies to the $\ket{eg}$ pair states (circles) as well as the $\ket{e\uparrow}$ and $\ket{e\downarrow}$ single-particle states (gray diamonds), compared to the zero-field single-particle transition (gray dashed line).
    Solid blue-red lines correspond to a fit of $E_\pm(B)$ from which we extract the corresponding scattering lengths.
    The color gradient reflects the overlap $|c_{\pm}|^2=|\braket{eg^\pm}{\psi}|^2$ of the eigenstate $\ket{\psi}$ with $\ket{eg^-}$ (upper branch) respectively $\ket{eg^+}$ (lower branch).
    The solid gray lines represent the differential Zeeman shift~\cite{SM}.
    All data points are extracted from Lorentzian fits to spectroscopic measurements described in the main text.
    (b)~Clock-line spectroscopy at $1.1\Gauss$ of atom pairs in $\ket{eg^-}$.
	The solid line corresponds to a double Lorentzian fit, showing a loss feature at $-5.1(1)\kHz$ corresponding to the formation of $\ket{ee}$ pairs.
    (c)~Lifetime of the $\ket{ee}$ state measured in a $31.2(8)\,E_\mathrm{rec}$ deep single-axis optical lattice.
    We show the number $n(t)$ of remaining $\ket{e}$ atoms after varying hold time $t$ in a spin-polarized (black circles) and spin-balanced sample (green squares).
    The solid lines denote fits to the data with an exponential decay (black) or a two-body decay $n(t)=n(0)/\left[1+n(0)\beta_{ee}t\right]$ (green)~\cite{SM}.
  }
\end{figure}

We determine these scattering lengths with a spectroscopy technique similar to the one shown on Fig.~\ref{fig:feshbach}.
Here, we focus on low magnetic fields $\leq 50\Gauss$ and measure the energy branches corresponding to $\ket{eg^+}$ as well as $\ket{eg^-}$, the latter connecting to the $\ket{o}$ branch in Fig.~\subfigref{feshbach}{c} at large magnetic fields.
We apply high-resolution coherent clock pulses to couple the initial $\ket{gg}$ state to the desired state.
In the spectra shown in Fig.~\subfigref{int_spectroscopy}{a}, we identify two branches corresponding to the energies $E_\pm(B)-U_{gg}$ with respect to the single-particle transition energy at zero magnetic field.
Here, $U_{gg}$ is the magnetic-field-independent interaction of the initial $\ket{gg}$ pair.
The magnetic field dependence of the energy branches at low fields is induced by the differential Zeeman shift $\delta(B)$ via $E_{\pm}(B)=V\pm\sqrt{V_{ex}^2+\delta(B)^2}$~\cite{scazza:2014}.
Here, $V = {\left(U_{eg}^+ + U_{eg}^-\right)}/2$ and $V_{ex} = {\left(U_{eg}^+ - U_{eg}^-\right)}/2$ denote the direct and spin-exchange energy, respectively.
We adjust the Rabi frequency to compensate for the magnetic-field-dependent super- and subradiance of the pair states~\cite{SM}.
From these energies, we extract $a_{eg}^+$ as well as $a_{eg}^-$ using a fit to a similar interaction model as for the OFR but with individual constant scattering lengths~\cite{busch:1998, SM}.

We find $a_{eg}^+= 240(4)\,a_0$ and $a_{eg}^-= 389(4)\,a_0$, where the initial state interaction energy is determined by $a_{gg}=-2.8(3.6)\,a_0$~\cite{kitagawa:2008}.
The error is dominated by the uncertainty in $a_{gg}$ since the relative quantities $a_{eg}^+ - a_{gg}= 242.7(1)\,a_0$ and $a_{eg}^- - a_{gg}= 392.2(2)\,a_0$ show an uncertainty smaller by an order of magnitude~\cite{SM}.
Crucially, these results imply that $V_{ex} < 0$, corresponding to antiferromagnetic interorbital spin-exchange, in qualitative agreement with a similar measurement~\cite{ono:2019}. 
The $p$-wave interaction in higher temperature $\yb{171}$ gases has been previously analyzed in Ref.~\cite{lemke:2011}.
The presence of both ferromagnetic and antiferromagnetic exchange in $\yb{173}$~\cite{scazza:2014,cappellini:2014} and $\yb{171}$, respectively, therefore enables the implementation of two-orbital Hamiltonians with either type of interaction.

%
%
To complete the characterization of all interaction channels, we measure the scattering length $a_{ee}$ associated to a $\ket{ee}$ pair, which has not yet been determined in \yb{171}.
With two consecutive excitation pulses, we drive atom pairs to $\ket{eg^-}$ and subsequently to $\ket{ee}$.
When the second pulse is on resonance, we observe a loss of $\ket{e}$ atoms, as visible in Fig.~\subfigref{int_spectroscopy}{b}.
This is due to the short lifetime of on-site $\ket{ee}$ pair states, which is expected~\cite{gorshkov:2010} and suggested by the broad loss feature.
We extract $a_{ee}=104(7)\,a_0$ from the spectral line using the same model as for the interorbital scattering lengths.

We additionally characterize the strongly inelastic collisions between two $\ket{e}$ atoms in a single-axis optical lattice at $30\Gauss$.
Here, the confinement is reduced in order to decrease the density and make the experimental determination of the lifetime feasible.
We employ two consecutive clock pulses much shorter than the $\ket{ee}$ lifetime addressing the single-particle transitions $\ket{g\downarrow}\rightarrow \ket{e\downarrow}$ and $\ket{g\uparrow}\rightarrow \ket{e\uparrow}$ to prepare the initial state and monitor the number of remaining $\ket{e}$ atoms for varying hold times [see Fig.~\subfigref{int_spectroscopy}{c}].
We find particularly short lifetimes and estimate a large effective two-body loss coefficient $\beta_{ee} = 4.8(2.1)\times10^{-12}\asciimathunit{cm^3/s}$~\cite{SM}, which is comparable to a previous measurement with non-degenerate atoms~\cite{ludlow:2011}.
The fast decay of the $\ket{ee}$ state highlights the importance to localize individual $\ket{e}$ atoms on lattice sites in quantum simulation experiments, which can be achieved with state-dependent optical lattices~\cite{riegger:2018}.

\begin{figure}[t!]
  \includegraphics[width=\columnwidth]{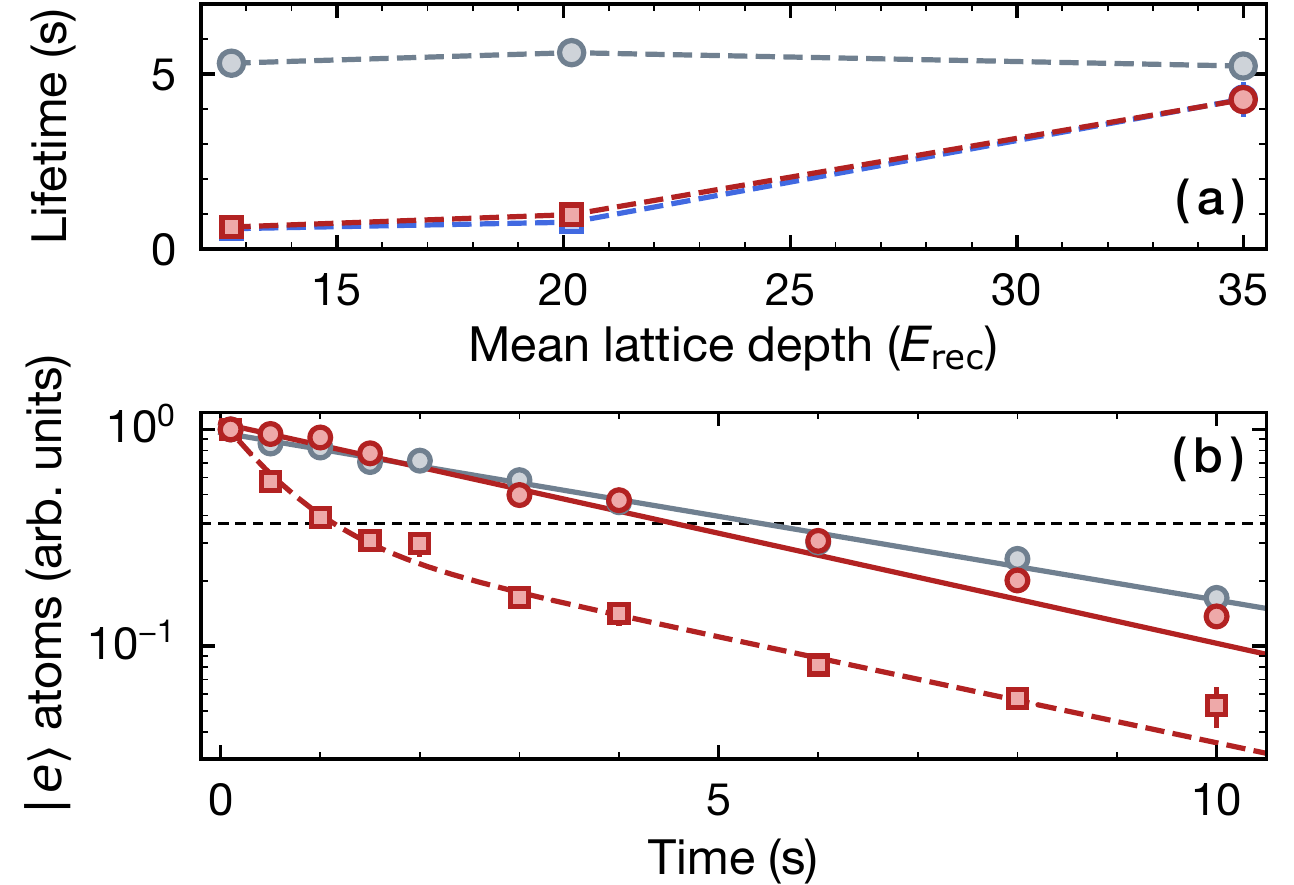}
  \caption{\label{fig:lifetimes}
    Lifetime of the pair states in a 3D optical lattice.
    (a)~Lifetime of $\ket{eg^+}$ (red) and $\ket{eg^-}$ (blue) states, compared to the lifetime of spin-polarized $\ket{e}$ atoms (gray).
    The dashed lines serve as a guide to the eye.
    We fit a sum of two exponentials (squares) or a single exponential (circles) to the atom number time traces.
    (b)~Sample data and fits at $34.99(6)\,E_\mathrm{rec}$ (circles) respectively $20.18(5)\,E_\mathrm{rec}$ (squares) with sum of two exponentials (red dashed line) or single exponential fits (solid lines).
    The dashed horizontal line indicates the $1/e$ remaining fraction used to define the lifetime of the states.
  }
\end{figure}

%
%
Finally, we probe the lifetimes of the interorbital pair states in a 3D optical lattice at $25\Gauss$.
Long lifetimes of these states are essential for the study of multiorbital many-body physics.
We prepare the pair states with high-intensity clock pulses and measure the decay with a second pulse after a variable hold time.
In Fig.~\subfigref{lifetimes}{a}, we show the corresponding fits for different lattice depths and compare with the decay of $\ket{e\uparrow}$ atoms prepared in a spin-polarized sample.
While the latter depends only weakly on the confinement, we find that the $\ket{eg^\pm}$ lifetimes are significantly reduced in shallower lattices.
Since the $\ket{g}$ atom number remains constant, we assume the losses are caused by residual tunnelling of $\ket{e}$ atoms.
In a $35\,\Erec$ deep lattice, the observed lifetimes correspond to an inelastic loss-rate coefficient of $\beta_{eg}^\pm\leq 2.6(3)\times 10^{-16}\asciimathunit{cm^3/s}$~\cite{SM}.
We only specify an upper bound for $\beta_{eg}^\pm$ since our value also includes single-particle losses and effects of tunneling.
Overall, the large observed $\ket{eg^\pm}$ lifetimes, longer than those reported in the high temperature regime \cite{ludlow:2011}, are very suitable for implementations of two-orbital Hamiltonians with \yb{171} in state-dependent optical lattices.

%
%
In conclusion, we have directly produced and characterized deeply-bound dimers in \yb{171} by addressing the narrow free-to-bound transition in a 3D optical lattice.
Due to their large binding energy at zero magnetic field, these are only weakly perturbed by the external confinement, in stark contrast to \yb{173}.
Moreover, we have determined the wavelength which makes the transition to the dimer state insensitive to the trap depth, providing a starting point for the state preparation in future molecular clock experiments.
The OFR, which we find at a magnetic field about $1300\Gauss$, makes it an interesting system for exploring the rich physics of strongly interacting multiorbital quantum gases such as in the case of Fermi polarons~\cite{darkwahoppong:2019}.
Furthermore, we have precisely measured the interorbital scattering lengths, which imply antiferromagnetic spin-exchange interactions, in qualitative agreement with Ref.~\cite{ono:2019}.
We also find low loss rates for interorbital atom pairs, making \yb{171} a promising platform for quantum simulations, in particular for Kondo impurity physics~\cite{kanasz-nagy:2018, nakagawa:2018, goto:2019, zhang:2020} and the Kondo lattice model~\cite{gorshkov:2010, foss-feig:2010, foss-feig:2010b, silva-valencia:2012}.




\begin{acknowledgments}
  We thank Jesper Levinsen and Meera M. Parish for insightful discussions.
  We also thank Florian Fertig and Caroline Tornow for technical contributions as well as the group of Gerhard Rempe for lending us equipment.
  This work was supported by the European Research Council through the synergy grant UQUAM and by the European Union’s Horizon 2020 funding.
  N.D.O. acknowledges funding from the International Max Planck Research School for Quantum Science and Technology.
\end{acknowledgments}


\bibliography{references}


\end{document}


\title{{\small Supplemental Material}\\Clock-line photoassociation of strongly bound dimers in a magic-wavelength lattice}

\author{O.~Bettermann}
\email{Oscar.Bettermann@physik.uni-muenchen.de}
\author{N.~\surname{Darkwah~Oppong}}
\thanks{These authors contributed equally to this work}
\author{G.~Pasqualetti}
\thanks{These authors contributed equally to this work}
\author{L.~Riegger}
\author{I.~Bloch}
\author{S.~F{\"o}lling}
\affiliation{Ludwig-Maximilians-Universit{\"a}t, Schellingstra{\ss}e 4, 80799 M{\"u}nchen, Germany}
\affiliation{Max-Planck-Institut f{\"u}r Quantenoptik, Hans-Kopfermann-Stra{\ss}e 1, 85748 Garching, Germany}
\affiliation{Munich Center for Quantum Science and Technology (MCQST), Schellingstra{\ss}e 4, 80799 M{\"u}nchen, Germany}
\date{\today}

\date{\today}

\maketitle

\renewcommand{\thefigure}{S\arabic{figure}}
\setcounter{figure}{0}
\renewcommand{\theequation}{S.\arabic{equation}}
\setcounter{equation}{0}
\renewcommand{\thesection}{S.\Roman{section}}
\setcounter{section}{0}
\renewcommand{\thetable}{S\arabic{table}}
\setcounter{table}{0}

\onecolumngrid

%
%


\subsection{State preparation}\label{sec:state_preparation}

Given the very small ground-state s-wave scattering length in \yb{171}, i.e., $a_{gg}=-2.8(3.6)\,a_0$~\cite{kitagawa:2008}, where $a_0$ denotes the Bohr radius, forced evaporative cooling has to be performed sympathetically with another isotope to ensure sufficiently fast thermalization of the system during the evaporation.
A natural choice for this is \yb{174}, owing to its large natural abundance~\cite{deLaeter:2006} and good interisotope s-wave scattering length with \yb{171}, i.e., $429(13)\,a_0$~\cite{kitagawa:2008}.
Both isotopes are consecutively loaded into a three-dimensional (3D) magneto-optical trap (MOT) working on the $\sS{0}\rightarrow\tP{1}$ intercombination line.
In order to trap both isotopes in the same MOT with a single laser, we make the latter resonant with the transition of \yb{171} and use an electro-optical modulator to generate a frequency sideband resonant with the corresponding transition of $\yb{174}$~\cite{pasqualetti:2018}.
The atoms are then loaded into a far-off-resonant crossed-beam optical dipole trap generated by a high-power $1064\nm$ laser.
After evaporative cooling, we typically have $\approx3\times10^4$ \yb{171} and $<10^2$ \yb{174} atoms in the dipole trap.
Subsequently, the atoms are adiabatically loaded into a deep 3D isotropic optical lattice operating at the so-called magic wavelength $\lambda_m=759.36\nm$~\cite{lemke:2009}.
At this particular wavelength, the atoms in the electronic \sS{0} ground state (denoted by $\ket{g}$) experience the same trapping potential as the atoms in the long-lived \tP{0} excited state (denoted by $\ket{e}$).
This prevents broadening of the $\ket{g}\rightarrow\ket{e}$ clock transition caused by the inhomogeneity of the trapping potential~\cite{katori:2003}.
By means of clock-line spectroscopy, we verify that all atoms occupy the lowest energy band in the lattice.
We compare the spectral amplitude of the pair and single-particle states to estimate that $\approx25{\%}$ of the atoms are on doubly-occupied lattice sites.



\subsection{Strong saturation absorption imaging}\label{sec:high_int_calib}

The column density of our atomic cloud is measured by means of absorption imaging.
The atoms in $\ket{g}$ are first imaged using the broad dipole-allowed $\sS0\rightarrow\sP1$ transition.
The atoms in $\ket{e}$ are subsequently repumped to $\ket{g}$ via the $\tP0\rightarrow\tD1$ transition with a measured efficiency of $70.5(5.1)\%$ and imaged on the $\sS0\rightarrow\sP1$ transition.
This allows to independently measure the population in $\ket{g}$ and $\ket{e}$.
The optical depth $\mathrm{od}(i,j)$ of the atomic cloud on a pixel of position $(i, j)$ on our camera is reliably extracted when performing strong-saturation absorption imaging by using the modified Beer-Lambert law~\cite{reinaudi:2007}
\begin{equation}
\mathrm{od}(i, j) = \alpha\ln\left[\frac{I_\mathrm{i}(i, j)}{I_\mathrm{f}(i, j)}\right]+\frac{I_\mathrm{i}(i, j)-I_\mathrm{f}(i, j)}{I_\mathrm{sat}},\label{eq:beer-lambert}
\end{equation}
where $I_\mathrm{sat}$ is the saturation intensity of the atomic transition and $I_\mathrm{f}$ ($I_\mathrm{i}$) denotes the imaging light intensity in the presence (absence) of atoms.
The parameter $\alpha$ accounts for the polarization of the imaging beam and the multi-level structure of the atom and has to be determined experimentally.
To determine the optimal value for $\alpha$, we measure $\mathrm{od}(i, j)$ for variable imaging light intensities.
The parameter $\alpha$ is then chosen to minimize the standard deviation of the obtained densities, such that the measured atom number is independent of the intensity of the imaging light~\cite{reinaudi:2007}.



\subsection{Optical lattice depth calibration}\label{sec:lattice_calib}

\begin{figure}
  \centering
	\includegraphics[width=\textwidth]{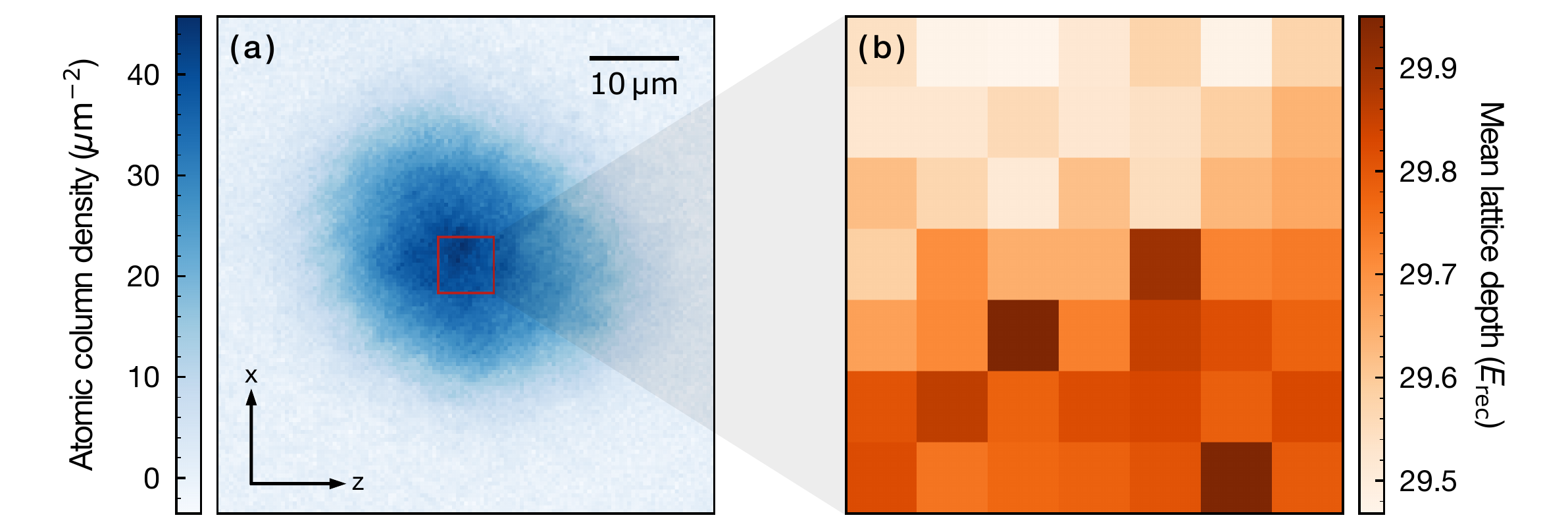}
  \caption{Spatially-resolved lattice depth calibration. 
  (a) Atomic column density measured via absorption imaging in a 3D optical lattice.
  The central red square defines the $6.58\um\times6.58\um$ ($14\times14$ pixels) region of interest (ROI) in which the lattice is calibrated.
  (b) Mean lattice depth inside the ROI measured via clock-line sideband spectroscopy, in units of the lattice photon recoil energy $\Erec$.
  We show the value obtained in each $0.94\um\times0.94\um$ ($2\times2$ pixels) square inside the ROI.
}\label{fig:lattice_calibration}
\end{figure}

In our experiments, we typically only consider a small region of interest (ROI) in the center of our absorption images, as illustrated in Fig.~\subfigref{lattice_calibration}{a}.
%
%
%
On the one hand, this improves the signal-to-noise ratio since the fraction of atoms on doubly-occupied lattice sites is larger in the center of the cloud.
On the other hand, it also reduces the effect of the spatial inhomogeneity of the lattice potential, which in particular broadens the $\ket{gg}\rightarrow\ket{eg^\pm}$ lines and thereby reduces the measurement accuracy of the on-site interaction energies $U_{eg}^\pm$.

Each lattice axis $i=x,y,z$ is characterized by driving atoms of a spin-polarized sample into the first excited energy band with a clock-laser pulse propagating along $i$.
The corresponding band gap yields the lattice depth $V_i$ along that axis.
The mean lattice depth $\bar{V}=(V_xV_yV_z)^{1/3}$ is obtained by combining the measurements for all axes.
In Fig.~\subfigref{lattice_calibration}{b}, we show $\bar{V}$ as a function of the position in the center of a typical absorption image.
The spatial inhomogeneity of the lattice potential can clearly be seen, thereby highlighting the relevance of our calibration method.
The lattice is calibrated before and after each measurement in order to keep track of any drifts.



\subsection{Differential and quadratic Zeeman shifts}\label{sec:zeeman_shifts}
Since $\ket{g}$ and $\ket{e}$ are distinct electronic states, their hyperfine states ($m_F = \pm 1/2$) also possess different Land{\'e} g-factors.
As a consequence, the frequency of the electronic transition from an initial state $\ket{g, m_F}$ to a final state $\ket{e,m_{F'}}$ is modified to first order under the effect of an external magnetic field $B$ by a linear Zeeman shift
\begin{equation}
\Delta\nu\left(B, m_F\rightarrow m_{F'}\right)=\frac{\mu_B B}{2\pi\hbar}\left[\left(m_{F}-m_{F'}\right)g_I+m_{F'}\delta g\right],\label{eq:transition_freq}
\end{equation}
where $\mu_B$ is the Bohr magneton and $\delta g$ defines the differential Land{\'e} g-factor between $\ket{e}$ and $\ket{g}$.
The nuclear g-factor $g_I$ of \yb{171} is given by $g_I = \mu\mu_N/|I|\mu_B$, where $\mu_N$ is the nuclear magneton, $I=1/2$ is the nuclear spin quantum number and $\mu=0.4919\,\mu_N$ is the nuclear magnetic moment~\cite{sansonetti:2005}.
Since both electronic states have a total angular momentum $F=1/2$, there are only four different transitions, and we define for simplicity
\begin{align*} 
\Delta\nu_1(B) \equiv \Delta\nu(B, +1/2\rightarrow +1/2), \quad
\Delta\nu_2(B) \equiv \Delta\nu(B, -1/2\rightarrow -1/2), \\
\Delta\nu_3(B) \equiv \Delta\nu(B, +1/2\rightarrow -1/2),\quad
\Delta\nu_4(B) \equiv \Delta\nu(B, -1/2\rightarrow +1/2).
\end{align*}
In our experiment, in order to compensate long-term drifts of the clock laser and additional non-linear shifts such as the second-order Zeeman shift,  all four transition frequencies are measured and shifted according to $\Delta\nu_{1,2}\rightarrow\Delta\nu_{1,2}-(\Delta\nu_1+\Delta\nu_2)/2$ and $\Delta\nu_{3,4}\rightarrow\Delta\nu_{3,4}-(\Delta\nu_3+\Delta\nu_4)/2$.
Combining all transition frequencies, the value of the magnetic field can be calibrated via~\cite{boyd:2007, ludlow:2015}
\begin{equation}
B = \frac{\pi\hbar\left[\Delta\nu_1(B)+\Delta\nu_3(B)-\Delta\nu_2(B)-\Delta\nu_4(B)\right]}{\mu_Bg_I},\label{eq:b_field_self_calib}
\end{equation}
leading to
\begin{equation}
\delta g=\frac{2g_I}{1+\displaystyle{\left[\frac{\Delta\nu_3(B)-\Delta\nu_4(B)}{\Delta\nu_1(B)-\Delta\nu_2(B)}\right]}}.\label{eq:diff_lande}
\end{equation}
This method has been successfully applied to \sr{87}~\cite{boyd:2007}, \yb{171}~\cite{lemke:2012} and \yb{173}~\cite{darkwahoppong:2019} and is independent of any prior magnetic field calibration.

In addition to the first-order differential Zeeman shift, a quadratic contribution $\Delta^{\left(2\right)}\nu\left(B\right) = \delta^{(2)}B^2$ from the second-order Zeeman shift also has to be accounted for, and is computed from
\begin{equation}
\delta^{(2)} = \frac{\Delta\nu_1(B)+\Delta\nu_2(B)}{2B^2}=\frac{\Delta\nu_3(B)+\Delta\nu_4(B)}{2B^2},\label{eq:quadratic_zeeman}
\end{equation}
using Eq.~\eqref{eq:b_field_self_calib} as calibration for the magnetic field, similar to the calculation of the first-order Zeeman shift.

\begin{figure}
  \centering
	\includegraphics[width=\textwidth]{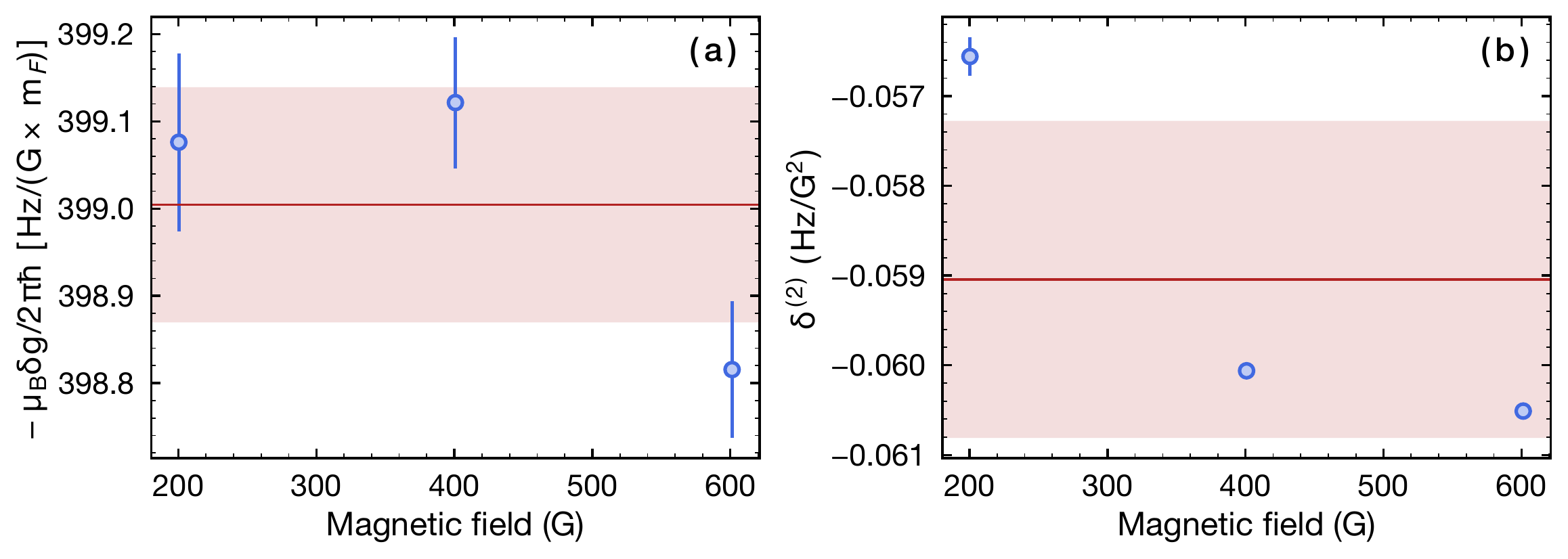}
  \caption{Characterization of the (a) first- and (b) second-order Zeeman shift via clock-line spectroscopy.
  For each magnetic field, we extract all transition frequencies $\Delta\nu_i$ defined in the main text via Loretzian fits to the spectroscopic data, and use Eq.~\eqref{eq:diff_lande} and \eqref{eq:quadratic_zeeman} to compute the values shown in (a) and (b), respectively.   
  The error bars are obtained considering the uncertainties on $\Delta\nu_i$ from the fit.
  The red line indicates the mean of the three data points, while the red area corresponds to their standard deviation.
  }\label{fig:zeeman_shifts}
\end{figure}

By means of clock-line spectroscopy at large magnetic fields in spin-polarized samples, we measure all four transition frequencies $\Delta\nu_i$ for $B=(200,\,400,\,600)$\Gauss.
Using Eq.~\eqref{eq:diff_lande}, we determine $\delta g$ for each magnetic field and obtain a mean value of $\delta g= -(2\pi\hbar/\mu_B)\times 399.0(1)\asciimathunit{Hz/(G}\, m_F)$, as shown on figure \subfigref{zeeman_shifts}{a}.
%
%
%
The error is defined as the standard deviation of the measured $\delta g$.
Similarly, we use Eq.~\eqref{eq:quadratic_zeeman} to extract $\delta^{(2)}=-0.059(2)\asciimathunit{Hz/G^2}$.
Overall, our results are in good agreement with previous measurements~\cite{lemke:2012, ono:2019}.



\subsection{Interaction energies around the orbital Feshbach resonance}
\label{sec:feshbach_model}

We characterize the orbital Feshbach resonance (OFR) by measuring the energy of the associated closed-channel molecular bound state as well as the energy of the repulsively interacting open-channel pair state.
To model the data, we assume the on-site trapping potential to be isotropic and harmonic, with a trapping frequency $\omega=\sqrt{\bar{V}/\Erec}\left(\Erec/\hbar\right)$, where $\bar{V}$ is the mean lattice depth and $\Erec$ the recoil energy of a lattice photon.
Furthermore, the interaction between two atoms is given by the regularized pseudopotential $V(\mathbf{r})=(4\pi\hbar^2a/m)\delta^{(3)}(\mathbf{r})(\partial/\partial r) r$, where $m$ is the atomic mass and $a$ the s-wave scattering length.
In this case, the problem separates into center-of-mass and relative coordinates and can be exactly solved~\cite{busch:1998}.
The eigenenergies $E$ of the system are the solutions of
\begin{equation}
\sqrt{2}\frac{\Gamma\left(-E/2\hbar\omega+3/4\right)}{\Gamma\left(-E/2\hbar\omega+1/4\right)}=\frac{a_{ho}}{a},\label{eq:busch_model}
\end{equation}
where $a_{ho}=\sqrt{\hbar/m\omega}$ is the harmonic oscillator length.
We take into account the anharmonicity of the lattice by means of first-order perturbation theory corrections to the energies computed via Eq.~\eqref{eq:busch_model}, expanding the lattice potential up to 8\textsuperscript{th} order.
Around the OFR, the behaviour can be described by an effective open-channel scattering length given by the simple expression~\cite{chin:2010}
\begin{equation}
a(B) = a_{bg}\left(1-\frac{\Delta}{B-B_0}\right),
\end{equation}
where $B_0$ is the resonance position, $\Delta$ is the resonance width and $a_{bg}$ is the background scattering length.
We fit this model to our data only around the resonance, where the bound state is assumed to be in the universal regime.
The corresponding magnetic field range is chosen such that it lies within a range $2\Delta$ around the resonance.

We note that the two-channel model that has been successfully applied to the OFR in $\yb{173}$~\cite{hoefer:2015} is not well-suited to describe our experimental data, even when considering analytically calculated effective ranges~\cite{flambaum:1999}.
We attribute this to the much larger linear Zeeman shift $\delta(B)\simeq 0.25\MHz$ in the vicinity of the resonance, which makes the scattering amplitude particularly sensitive to finite-range effects.



\subsection{Interorbital dimers in two dimensions}
\label{sec:bound_state_2D}

\begin{figure}
  \centering
	\includegraphics[width=\textwidth]{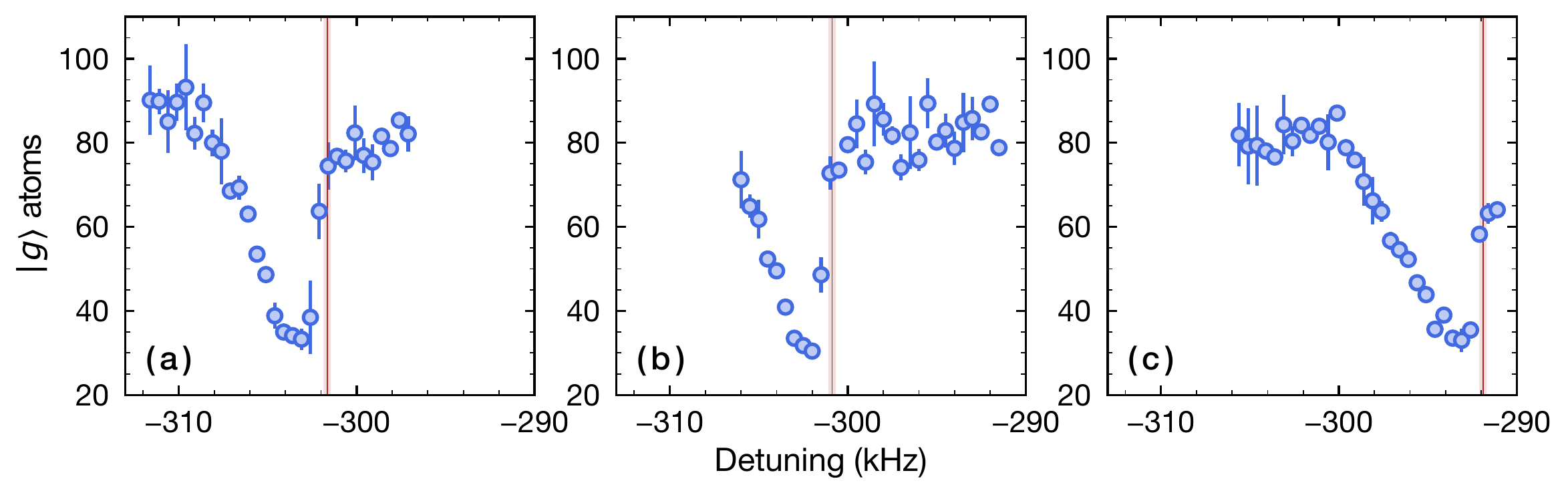}
  \caption{
  Clock-line photoassociation of free-space interorbital dimers in two-dimensional pancake-shaped traps formed by a $25.2(1)\,\Erec$ deep single-axis optical lattice at (a) $1.1\Gauss$, (b) $5\Gauss$, and (c) $50\Gauss$.
  The error bars correspond to the standard error of two consecutive measurements.
  The red vertical line denotes the expected position of the zero-momentum transition computed from Eq.~\eqref{eq:binding_energy_2D}, while the red area corresponds to the estimated uncertainty.
  }\label{fig:bound_state_2D}
\end{figure}

In addition to the on-site clock-line photoassociation of interorbital dimers, we similarly demonstrate their production in free space, in two-dimensional pancake-shaped traps.
We load the atoms in a single-axis magic-wavelength optical lattice, with an additional confinement provided by a crossed dipole trap.
Colliding atomic pairs are then driven at different magnetic fields $B$ to the dimer state with a clock-laser pulse propagating along the lattice axis $z$.
For $B\lesssim200\Gauss$, the dimer binding energy $\epsilon_b=h\times 292.1(2)\kHz$ is almost constant and the zero-momentum dimer transition frequency $E(B)/h$ is derived from the binding energy as
\begin{equation}
E(B)= -\epsilon_b - \epsilon_{0, z} - \hbar (\omega_x + \omega_y)/2 - \delta(B)\label{eq:binding_energy_2D}
\end{equation}
relative to the $\ket{g\uparrow}\rightarrow\ket{e\uparrow}$ transition.
Here, $\epsilon_{0,z}=h\times 9.62(2)\kHz$ is the energy of the lattice ground band, $\omega_x = 2\pi \times 53.7(8.1)\Hz$ and $\omega_y = 2\pi\times 262.6(2.8)\Hz$ are the trapping frequencies along the weakly confined axes $x$ and $y$, and $\delta(B)=-h\times 399.0(1)\asciimathunit{(Hz/G)}\times B/2$ is the differential Zeeman shift of the single-particle transition.
The above equation takes into account the difference in zero-point energy of the initial two-particle and final molecular states.
We find good agreement with our experimental data, as can be seen in Fig.~\ref{fig:bound_state_2D}.
%
%
%
The asymmetric lineshape is a consequence of the continuous momentum distribution in the sample.



\subsection{Inter- and intraorbital scattering lengths}
\label{sec:scattering_lengths}

The on-site interorbital interaction energies $U_{eg}^\pm$ are measured by coherently driving $\ket{gg}$ pairs into the interorbital states (corresponding to $\ket{eg^+}$ and $\ket{eg^-}$ at zero magnetic field) with linearly polarized $\pi$-pulses at low magnetic fields ($B\leq 50$\Gauss).
We additionally measure the frequency of the single-particle transitions $\ket{g\downarrow}\rightarrow\ket{e\downarrow}$ and $\ket{g\uparrow}\rightarrow\ket{e\uparrow}$ as absolute frequency reference.
Long-term drifts of the clock laser and contributions from the quadratic Zeeman shift are suppressed by setting the mean of both single-particle transition frequencies to zero.

The Rabi frequencies $\Omega_{1,2}$ of the transitions between $\ket{gg}$ and both interorbital pair states differ from the single-particle Rabi frequency $\Omega_0$ and depend on the magnetic field.
Starting from a generic Hamiltonian including contributions from on-site interactions, the differential Zeeman shift and atom-light interactions, one can show that~\cite{scazza:2015}
\begin{equation}
\displaystyle{\frac{\Omega_{1,2}(B)}{\Omega_0}=\frac{\left(\mp1+\sqrt{1+\frac{\delta(B)^2}{V_{\mathrm{ex}}^2}}\right)}{\sqrt{1+\frac{\delta(B)^2}{V_{\mathrm{ex}}^2}}}\sqrt{1+\frac{V_{\mathrm{ex}}^2}{\delta(B)^2}\left(1\pm\sqrt{1+\frac{\delta(B)^2}{V_{\mathrm{ex}}^2}}\right)}}, \label{rabi_couplings}
\end{equation}
where $V_{\mathrm{ex}}=(U_{eg}^+-U_{eg}^-)/2$ is the spin-exchange interaction energy and $\delta(B)$ is the differential Zeeman shift.
At zero magnetic field, the coupling to $\ket{eg^-}$ is enhanced by a factor of $\sqrt{2}$ with respect to the single particle transition, while the coupling to $\ket{eg^+}$ vanishes.
At finite magnetic fields, however, the mixing between $\ket{eg^+}$ and $\ket{eg^-}$ induces a finite coupling to both interorbital pair states, yielding $\Omega_{1,2}/\Omega_0\rightarrow 1$ in the limit of large fields.
This change in Rabi frequency is compensated by modifying the intensity of the excitation pulses accordingly, ensuring a constant Fourier broadening of the spectral lineshapes throughout the whole measurement with the exception of the 1.1\Gauss~and 2.5\Gauss~data, where the length of the pulse addressing $\ket{gg}\rightarrow\ket{eg^+}$ is increased.

From the known on-site interaction energies and lattice calibration, the corresponding s-wave scattering lengths can be determined.
A first approach is to consider two indistinguishable particles on a single 3D lattice site interacting via the pseudopotential $V(\mathbf{r})=(4\pi\hbar^2a/m)\delta^{(3)}(\mathbf{r})$, where $\mathbf{r}$ is the interparticle distance, $m$ is the atomic mass and $a$ is the s-wave scattering length.
The corresponding Hubbard on-site interaction energy $U$ is then given by
\begin{equation}
U = \frac{4\pi\hbar^2a}{m}\prod_{i=x,y,z}\left(\int\mathrm{d}r|\psi_i(r)|^4\right),\label{eq:int_energy}
\end{equation}
where $i$ denotes the lattice axis.
In a simple approximation, one can assume the wavefunctions $\psi_i(r)$ to be given by the Wannier function $W_0(r)$ of the lowest lattice band.
This is valid in the limit of deep lattices and small scattering lengths, where contributions from higher bands are negligible.
From this, which is the method used in Ref.~\cite{ono:2019}, we obtain $a_{eg}^+-a_{gg} = 248.2(1)\,a_0$ and $a_{eg}^--a_{gg} = 391.7(2)\,a_0$ as interorbital scattering lengths.
A second way to access $a$ which better accounts for contributions from higher bands is to use the exact energy solutions for two interacting ultracold atoms in a harmonic trap~\cite{busch:1998}.
Here, we match the slope of the obtained energy branch around $a=0$ with the slope from Eq.~\eqref{eq:int_energy} as correction for the lattice anharmonicity.
With this, we obtain $a_{eg}^+-a_{gg} = 242.7(1)\,a_0$ and $a_{eg}^--a_{gg} = 392.2(2)\,a_0$, which are the values reported in the main text.
An exact diagonalization calculation~\cite{cappellini:2014} applied to our data agrees within $\approx1\%$ with the values obtained from this second method.

\begin{figure}
  \centering
	\includegraphics[width=\textwidth]{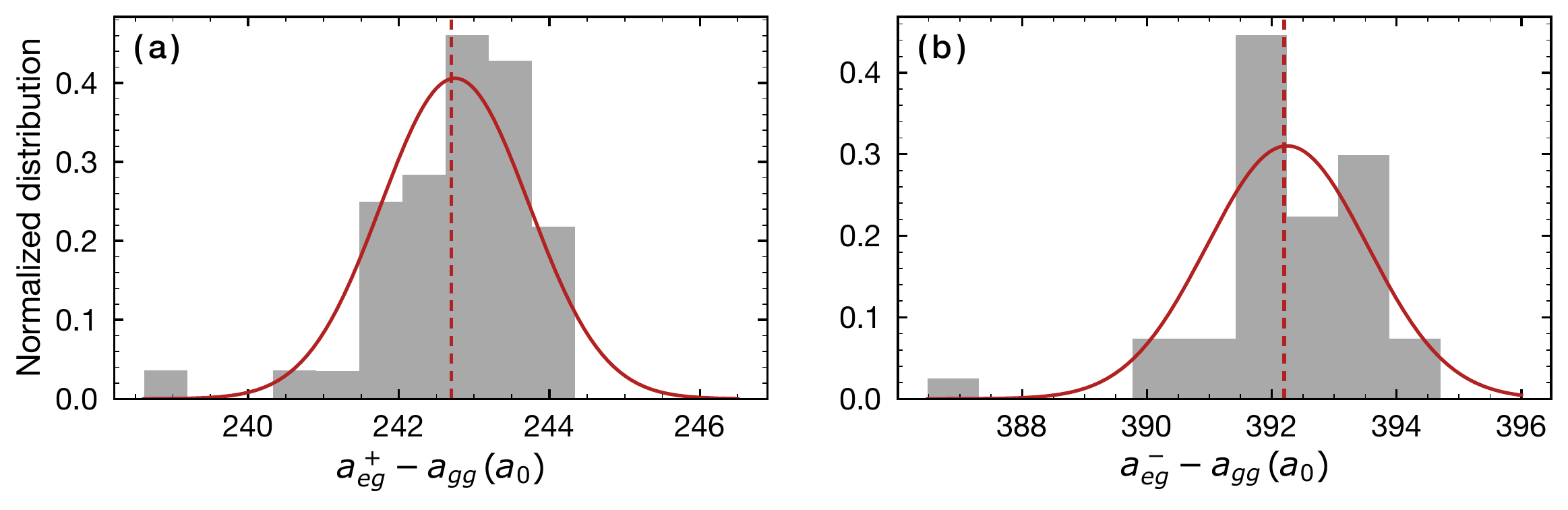}
  \caption{
	Normalized distribution of the interorbital scattering lengths $a_{eg}^\pm-a_{gg}$ measured in different areas of our absorption images, weighted by the atomic density in the corresponding area (grey bars).
	The red lines correspond to gaussian distributions with the mean value and the standard deviation of the data.
	Using the mean value of each distribution (dashed red lines), we find (a) $a_{eg}^+-a_{gg}=242.7(1)\,a_0$ and (b) $a_{eg}^--a_{gg}=392.2(2)\,a_0$, where $a_0$ is the Bohr radius.
	The error is given by the fit uncertainty on the position of the maximum of the distribution.
  }\label{fig:scattering_lengths}
\end{figure}

We extract the on-site interaction energies $U_{eg}^\pm$ in each $0.94\um\times0.94\um$ ($2\times2$ pixels) area shown in Fig.~\subfigref{lattice_calibration}{b}.
Using the corresponding lattice depth calibration leads to 49 different values for $a_{eg}^\pm$, as shown in Fig.~\ref{fig:scattering_lengths}.
%
%
%
The values reported in the main text correspond to the mean of the distribution, which we assume to be Gaussian, with an error given by the fit uncertainty on the position of the mean.
In table~\ref{tab:summary}, we summarize the values of all intra- and interorbital s-wave scattering lengths in \yb{171}, as well as the corresponding two-body loss rate coefficients (see below). 
%
%
\begin{table}[t]
  \caption{\label{tab:summary}%
    Summary of the $\sS{0}$--$\tP{0}$ s-wave scattering lengths and two-body loss coefficients in \yb{171}.}
  \begin{ruledtabular}
  \begin{tabular}{l c c c c c c}
  & \multicolumn{4}{c}{s-Wave scattering lengths ($a_0$)} & \multicolumn{2}{c}{Two-body loss coefficients $(\asciimathunit{cm^3/s})$} \\[-0.75em]
  & \multicolumn{4}{c}{\rule{180pt}{0.4pt}} & \multicolumn{2}{c}{\rule{185pt}{0.4pt}} \\[-0.3em]
   &$a_{gg}$ & $a_{eg}^+$ & $a_{eg}^-$ & $a_{ee}$ & $\beta_{eg}^\pm$ & $\beta_{ee}$ \\[0.45em] \hline
   Kitagawa et al.~\cite{kitagawa:2008} & $-2.8(3.6)$ & -- & -- & -- & -- & -- \\
   Ono et al.~\cite{ono:2019} & -- & $225(13)$ & $355(6)$ & -- & -- & -- \\   
   This work & -- & $240(4)$ & $389(4)$ & $104(7)$ & $\leq2.6(3)\times10^{-16}$ & $4.8(2.1)\times10^{-12}$   
    \end{tabular}
  \end{ruledtabular}
\end{table}
%
Similar measurements have also been previously performed in thermal samples~\cite{lemke:2011, ludlow:2011}.



\subsection{Two-body losses of clock-state atoms}\label{sec:ee_losses}

Two \yb{171} atoms in the $\ket{e}$ orbital and in distinct nuclear spin states are expected to interact via strongly inelastic collisions~\cite{gorshkov:2010}, leading to a rapid trap loss of these atoms.
Assuming a spin-unpolarized sample of $\ket{e}$ atoms characterized by a density $n_e(t)$ leads to the rate equation
\begin{equation}
\dot{n}_e(t) = -\beta_{ee}n_e^2(t),\label{eq:rate_eq}
\end{equation}
where $\beta_{ee}$ is the two-body loss rate coefficient between two $\ket{e}$ atoms~\cite{ludlow:2011}. The solution of Eq.~\eqref{eq:rate_eq} is
\begin{equation}
n_e(t)=\frac{n_0}{1+n_0\beta_{ee}t},\label{eq:density_profile}
\end{equation}
where $n_0$ is the initial density of the Fermi gas, i.e., $n_0=n_e(0)$.

We experimentally determine the coefficient $\beta_{ee}$ using a spin-balanced sample of $\ket{e}$ atoms in a single-axis optical lattice, with an additional confinement provided by a crossed dipole trap, similar to the measurement described in Ref.~\cite{scazza:2014}.
We track the number of atoms remaining in the trap as a function of time and fit the result to Eq.~\eqref{eq:density_profile}.

In order to compute the initial atomic density, we model the optical lattice as an array of quasi-two-dimensional systems.
In each layer of the lattice, we model the local density profile $n(x, y)$ as a two-dimensional Thomas-Fermi density distribution:
\begin{equation}
n(x,y) = \frac{mk_BT}{\pi\hbar^2}\ln\left\lbrace 1+z\exp\left[-\frac{m}{2k_BT}\left(\omega_x^2x^2+\omega_y^2y^2\right)\right]\right\rbrace,
\end{equation}
where $m$ is the atomic mass, $k_B$ is the Boltzmann constant, $\hbar$ is the reduced Planck constant, $T$ is the temperature of the gas, $z$ is the fugacity of the gas and $\omega_x$ and $\omega_y$ are the trapping frequencies along the weakly confined axes $x$ and $y$.
From center-of-mass oscillations after a sudden trap displacement, we determine $\omega_x = 2\pi \times 53.7(8.1)\Hz$ and $\omega_y = 2\pi \times 262.6(2.8)\Hz$. 
Considering only a small number of lattice layers on the order of our imaging resolution, we find the temperature $T=117(41)\,\mathrm{nK}$ from directly fitting the in-situ column integrated density $n(x)=\int \!\mathrm{d}y\,n(x, y)$.
We compute the density profile $n(z)$ along the (strongly confined) lattice axis by assuming the atoms to be in the vibrational ground state of the resulting harmonic potential:
\begin{equation}
n(z) = \sqrt{\frac{m\omega_z}{\pi\hbar}}\exp\left(-\frac{m\omega_z}{\hbar}z^2\right).
\end{equation}
From the lattice depth $V_z = 31.1(8) E_\mathrm{rec}$, we obtain the trapping frequency $\omega_z = 2 \sqrt{V_z/E_\mathrm{rec}} \left(E_\mathrm{rec} / \hbar\right) = 2\pi \times 22.6(3)\kHz$.
The total density is therefore given by $n(x,y,z)=n(x,y)n(z)$ and we find the average density
\begin{equation}
  \overline{n} = \frac{\int \!\mathrm{d}^3{\mathbf{r}}\; n(x, y, z)^2
}{\int \!\mathrm{d}^3{\mathbf{r}}\; n(x, y, z)} = 2.0(7)\times10^{13}\mathrm{cm}^{-3},
\end{equation}
which we use as initial density $n_0$ when fitting our experimental data to Eq.~\eqref{eq:density_profile}.



\subsection{Two-body losses in interorbital states}\label{sec:eg_losses}

Similar to the two-body losses of atoms in the $\ket{e}$ state, we characterize the losses of $\ket{eg^\pm}$ pairs in a deep 3D optical lattice. Assuming the atoms are in the ground band, the corresponding two-body loss rate coefficient $\beta_{eg}^\pm$ is related to the lifetime $\tau_{eg}^\pm$ associated to the exponential decay of the $\ket{eg}$ pairs via~\cite{garciaripoll:2009}
\begin{equation}
\left(\tau_{eg}^\pm\right)^{-1}=2\beta_{eg}^\pm\int \mathrm{d}^3r|W_0(\mathbf{r})|^4,\label{beta_eg}
\end{equation} 
where $W(\mathbf{r})$ is the ground-band Wannier function which determines the constant on-site density.
To estimate $\beta_{eg}^\pm$, we drive $\ket{gg}$ pairs to the interorbital states $\ket{eg^\pm}$ and measure the number of remaining atoms in the $\ket{e}$ as a function of time.
In a $35\Erec$ deep lattice, the data is well-described by an exponential decay and we compute $\beta_{eg}^\pm$ from Eq.~\eqref{beta_eg}, using the lifetime $\tau_{eg}^\pm$ fitted to our data.
We obtain the value reported in the main text, which is identical for both $\ket{eg^\pm}$ states.
Note that at this lattice depth, the lifetime of $\ket{eg}$ pairs is comparable to the lifetime of single $\ket{e}$ atoms.
We therefore overestimate $\beta_{eg}^\pm$ since our value also includes single-particle decay of $\ket{e}$ atoms and $\ket{ee}$ losses due to residual tunnelling of $\ket{e}$ atoms.  

\bibliography{references}